\documentclass[prl,twocolumn,aps,superscriptaddress]{revtex4-1}

\usepackage{epsfig}
\usepackage{graphicx}
\usepackage{amsmath}
\usepackage{amsfonts}

\begin{document}

\title{Emergent phenomena in multicomponent superconductivity: an introduction to the focus issue}

\author{Milorad V. Milo\v{s}evi\'c}
\affiliation{Departement Fysica, Universiteit Antwerpen, Groenenborgerlaan 171, B-2020 Antweren, Belgium}
\author{Andrea Perali}
\affiliation{School of Pharmacy, Physics Unit, University of Camerino, 62032 Camerino, Italy}

\begin{abstract}
Multicomponent superconductivity is a novel quantum phenomenon in many different superconducting materials, such as multiband ones in which different superconducting gaps open in different
Fermi surfaces, films engineered at the atomic scale to enter the quantum confined regime, multilayers, two-dimensional electron gases at the oxide interfaces, and complex materials in which different electronic orbitals or different carriers participate in the formation of the superconducting condensate. In all these systems the increased number of degrees of freedom of the multicomponent superconducting wave-function allows for emergent quantum effects that are otherwise unattainable in single-component superconductors. In this editorial paper we introduce the present focus issue, exploring the complex but fascinating physics of multicomponent superconductivity.
\end{abstract}

\maketitle

Very soon after the formulation of the Bardeen-Cooper-Schrieffer (BCS) theory, the prediction of multiband superconducting systems and the first extension of the BCS theory to two-band/two-gap superconductors were offered by Suhl, Matthias and Walker \cite{Suhl59}. The discovery of superconductivity in MgB$_2$ in 2001 
marked the formal appearance of the new class of superconductors: multiband/multigap ones, to which many of the recently discovered iron-based superconductors also belong. Multigap superconductivity arises when the gap amplitudes on different sheets of the Fermi surface are radically disparate, e.g. due to different dimensionality of the bands for the usual phonon-mediated pairing, as in the case of MgB$_2$, or due to the repulsive interband pairing interactions, as in the case of most iron-based superconductors, or due to the appearance of multiple Fermi-surface pockets dictated by the crystalline symmetry like in FeSe$_x$Te$_{1-x}$ and other unconventional multiband superconducting compounds. 

Multiband/gap superconductivity is thus emerging as a complex quantum coherent phenomenon with physical consequences which are different from or cannot be found at all in single-band superconductors. The cross-pairing between bands is energetically disfavored, hence multiple coupled condensates coexist and govern the overal superconducting behavior. The increased number of degrees of freedom allows for novel effects which are unattainable otherwise. Without claiming completeness, in what follows we point out some recent revelations based on multicomponent quantum physics in superconductors.

As mentioned above, the first two-gap superconductor with clearly distinct gaps in different bands was MgB$_2$ (for review, we refer to \cite{Eist,XXXi}). Even for this well-known material (studied for over a decade, with thousands of papers published), microscopic parameters and resulting superconducting length scales \cite{kogan,shan,komPRB} and magnetic behavior \cite{mosh15} are under ongoing debate. In that respect the behavior of vortex matter is crucial, since experimentally observed vortex states can serve as a smoking gun for the underlying physics. Cases where different condensates have very different electromagnetic properties, or coupling between them causes unconventional behavior (e.g. stemming from ‘frustration’ \cite{tesanovic,orlova}), can lead to novel vortex (nonuniform \cite{Guttierez,Chaves2011,Carl2011}, fractional \cite{geurts,pina,albino} and ‘skyrmion’ \cite{garaud,Gillis2014}) patterns - impossible otherwise.

Phase solitons \cite{Lin2012} and massive or massless Leggett modes \cite{Leggett1966,Marciani2013} are also possible benchmarks for multi-gap superconductivity, being associated with nontrivial phase differences between the condensates in different electronic bands, as detailed in Ref. \cite{TanakaSUST}. In that regime, the time-reversal symmetry breaking is possible, and can even survive moderate disorder (see Ref. \cite{StanevSUST}). Such states have not been observed to date, but could be experimentally accessible in multiband iron pnictides and chalcogenides. 

The latter iron-based materials form the other predominantly multi-band/gap family of superconductors (as probed by muon spin rotation \cite{KhasanovSUST}, spectroscopic measurements \cite{ZiemakSUST}, point-contact Andreev-reflection spectroscopy \cite{DagheroSUST}, magnetization and transport \cite{TamegaiSUST}, measurements of the penetration depth \cite{TanatarSUST}, etc.), but are much richer in novel physics than transition metal-borides. For one thing, the Cooper-pairing there is likely to be driven by electron-electron interactions, rather than the conventional electron-phonon coupling. Further, in iron-based superconductors experiments showed that weakly and strongly correlated conduction electrons coexist in much of the phase space, hence multiorbital physics becomes necessary to understand the phase diagram of these materials \cite{Capone2014}. Another interesting example is found in e.g. Ba$_{0.6}$K$_{0.4}$Fe$_{2}$As$_{2}$ (T$\rm _c$=37K), where two different $s$-wave gaps open in the different sheets of Fermi surface (FS): a large gap of $\Delta_2$=12 meV on the small FS and a small gap $\Delta_1$=6 meV in the large FS \cite{Ding2008}. The ratio $2\Delta_1/ \rm T_c$=3.7 is very close to the BCS value of 3.5, indicating BCS-like weakly coupled pairs in the large FS, while $2\Delta_2/ \rm T_c$=7.5 is very large and typical of BEC-like strongly coupled pairs in the small FS. Hence, the total superconducting condensate in Ba$_{0.6}$K$_{0.4}$Fe$_{2}$As$_{2}$ is a coherent mixture of BCS-like and BEC-like partial condensates (see also Ref. \cite{Okazaki}). Actually, various subgroups of iron-based superconductors show small Fermi surfaces at optimum doping where T$_c$ is the highest, appearing when the chemical potential is near a band edge, close to the bottom (if electron like) or top (if hole like) of the energy bands \cite{Borisenko2012}. In this situation, experiments show no evidences for nesting topology and the mechanism for high-T$_{\rm c}$ can be associated with the shape resonance scenario \cite{BianconiNP2013}. We refer to Ref. \cite{BianconiNP2013} for the schematic representation of the topology of Fermi surfaces for different superconducting iron-based materials, showing that in all cases large Fermi surfaces coexist with small Fermi surface pockets, supporting the two-band model for superconductivity as the minimal model to capture the band-edge physics and corresponding novel multi-band BCS-BEC crossover phenomena (see Ref. \cite{PeraliSUST}). This crossover regime has been recently detected by the collapse of the small Fermi surface pocket and electronic band dispersion becoming an inverted parabola in the coherent state \cite{Kanigel}. 

The multiband BCS-BEC crossover can thus determine the best situation for high-T$_c$ superconductivity, but also determine the optimal condition to allow the screening of superconducting fluctuations \cite{Perali2000}. Fluctuations in recent multiband materials are an interesting study object on their own (for example, Ref. \cite{BenfattoSUST} reports the peculiar differences between single-band and multi-band superconductors concerning the anisotropy of the fluctuations effects above and below T$_c$). As shown in Ref. \cite{VarlamovSUST}, fluctuations are very sensitive to the discrepancy in coherence lengths between the band-condensates, which in turn are most different in the presence of a shallow band (as discussed above) or in the vicinity of hidden criticality \cite{Komendova2012}, both of which are likely to be found in iron-based superconductors.

Interestingly, multiband superconductivity can also be induced by nanoengineering, even in elementary metals. Thanks to recent breakthroughs in nanofabrication, high-quality ultrathin films of Pb, Sn, In, Nb are now attainable, where multiband superconductivity is induced by confinement even though the bulk material is single-band (for brief review and theoretical challenges see Ref. \cite{ArkadySUST}). Namely, as first shown by Blatt and Thompson \cite{Blatt}, separate single-electron bands are formed due to quantum confinement, and can significantly shift in energy depending on the sample thickness. This leads to major changes of key superconducting properties each time when the bottom of a new single-electron band passes through the Fermi surface. Quantum confinement and shape resonances in nanoscale systems can induce sizeable T$_c$ amplifications \cite{Perali96,per2012}, experimentally confirmed in metallic nanowires of Al and Sn \cite{Shanenko2006}. Ultrathin high-quality metallic films are integrable in electronic circuits \cite{yam}, and will be highly responsive to applied electric field \cite{boz} - a desirable feature for any transistor for example. Furthermore, ultrathin multiband superconductors can also host the BCS-BEC crossover, in analogy to the above case in iron-based superconductors, and are strongly influenced by fluctuations - which crosslinks the physics of these two rather different superconducting systems. Note that on the level of fundamental physics, here discussed issues are very closely related to multi-subband superfluidity in confined ultracold Fermi gases (see Ref. \cite{Shanenko2012}) and electric-field-induced surface superconductivity in oxides \cite{Mizohata}.

In all above systems, understanding the underlying physics and the effects of hybridization between multiple quantum condensates in a single system, is clearly a pathway to yet unseen phenomena in fundamental science and practical achievements that can lead to future applications. This is the key objective of the current Focus Issue of Superconductor Science and Technology and the conference series `MultiSuper' (the first in this conference series was held in Lausanne (Switzerland) in 2012, with a sequel held in Camerino (Italy) in 2014 \cite{MultiSuper}). The scope of this action is broader than just multiband superconductors, and comprises other forms of {\it multicomponent} superconductivity, e.g. one found in artificial multicomponent coherent systems (e.g. multilayers \cite{AlbinoSUST}), multi-carrier superconductivity at oxide interfaces \cite{CapraraSUST}, or superconducting materials with multiple competing orders or symmetries of the order parameter \cite{MauroSUST, BianconiSUST}, a complexity certainly present in many copper-oxide, iron-pnictide, and heavy-fermion superconductors. All together, we hope that this Focus Issue will provide an illustrative snapshot of the current state of research in multicomponent superconductivity in a wide range of materials, and a roadmap for further investigations in this booming field. 

\section{Acknowledgments}
\begin{acknowledgments}
M.V.M. acknowledges support from the Research Foundation - Flanders (FWO) and the Special Research Funds of the University of Antwerp (BOF-UA). A.P. acknowledges financial support from the University of Camerino under the project FAR ``Control and enhancement of superconductivity by engineering materials at the nanoscale''. The authors thank the colleagues involved in the MultiSuper International Network (http://www.multisuper.org) for exchange of ideas and suggestions for this editorial.
\end{acknowledgments}



\begin{thebibliography}{99}
\bibitem{Suhl59}
Suhl H, Matthias B T and Walker L R 1959 {\em Phys. Rev. Lett} {\bf 3} 552

\bibitem{Eist}
Eisterer M 2007 {\em Supercond. Sci. Technol.} {\bf 20} R47

\bibitem{XXXi} Xi X X 2008 {\em Rep. Prog. Phys.} {\bf 71} 116501

\bibitem{kogan} Kogan V G and Schmalian J 2011 {\em Phys. Rev. B} {\bf 83} 054515

\bibitem{shan} Shanenko A A, Milo\v{s}evi\'c M V, Peeters F M and Vagov A V
2011 {\em Phys. Rev. Lett.} {\bf 106} 047005

\bibitem{komPRB} Komendov\'a L, Milo\v{s}evi\'c M V, Shanenko A A and
Peeters F M 2011 {\em Phys. Rev. B} {\bf 84} 064522

\bibitem{mosh15} Moshchalkov V, Menghini M, Nishio T, Chen Q H,
Silhanek A V, Dao V H, Chibotaru L F, Zhigadlo N D and
Karpinski J 2009 {\em Phys. Rev. Lett.} {\bf 102} 117001

\bibitem{tesanovic} Stanev V and Te\v{s}anovi\'c Z 2010 {\em Phys. Rev. B} {\bf 81} 134522

\bibitem{orlova} Orlova N V, Shanenko A A, Milo\v{s}evi\'c M V, Peeters F M,
Vagov A and Axt V M 2013 {\em Phys. Rev. B} {\bf 87} 134510

\bibitem{Guttierez} Gutierrez J, Raes B, Silhanek A V, Li L J, Zhigadlo N D, Karpinski J, Tempere J, and Moshchalkov V V 2012 {\em Phys. Rev. B} {\bf 85} 094511 

\bibitem{Chaves2011}
Chaves A, Komendov\'{a} L, Milo\v{s}evi\'{c} M V, Andrade J S Jr.,  Farias G A and Peeters F M 2011 {\em Phys. Rev. B} {\bf 83} 214523

\bibitem{Carl2011} Carlstr\"{o}m J, Garaud J and Babaev E 2011 {\em Phys. Rev. B} {\bf 84}
134518

\bibitem{geurts} Geurts R, Milo\v{s}evi\'c M V and Peeters F M 2010 {\em Phys. Rev. B}
{\bf 81} 214514

\bibitem{pina} Pi\~{n}a J C, de Souza Silva C C and Milo\v{s}evi\'c M V 2012 {\em Phys. Rev. B} {\bf 86} 024512

\bibitem{albino} da Silva R M, Milo\v{s}evi\'c M V, Dom\'inguez D, Peeters F M, and Aguiar J A 2014 {\em Appl. Phys. Lett.} {\bf 105} 232601

\bibitem{garaud} Garaud J, Carlstr\"{o}m J and Babaev E 2011 {\em Phys. Rev. Lett.} {\bf 107} 197001

\bibitem{Gillis2014}
Gillis S, J\"{a}ykk\"{a} J and Milo\v{s}evi\'{c} M V 2014 \textit{Phys. Rev. B} \textbf{89} 024512

\bibitem{Lin2012}
Lin S Z and Hu X 2012 \textit{New J. Phys.} \textbf{14} 063021

\bibitem{Leggett1966}
Leggett A J 1966 \textit{Prog. Theor. Phys.} \textbf{36} 901

\bibitem{Marciani2013}
Marciani M, Fanfarillo L, Castellani C and Benfatto L 2013
\textit{Phys. Rev. B} \textbf{88} 214508

\bibitem{TanakaSUST}
Tanaka Y 2015 {\em Supercond. Sci. Technol.} {\bf 28} 034002

\bibitem{StanevSUST}
Stanev V 2015 
{\em Supercond. Sci. Technol.} {\bf 28} 014006

\bibitem{KhasanovSUST}
Khasanov R and Guguchia Z
2015 {\em Supercond. Sci. Technol.} {\bf 28} 034003

\bibitem{ZiemakSUST}
Ziemak S, Kirshenbaum K, Saha S R, Hu R, Reid J-Ph, Gordon R, Taillefer L, Evtushinsky D, Thirupathaiah S, Büchner B, Borisenko S V, Ignatov A, Kolchmeyer D, Blumberg G and Paglione J 2015 
{\em Supercond. Sci. Technol.} {\bf 28} 014004

\bibitem{DagheroSUST}
Daghero D, Pecchio P, Ummarino G A, Nabeshima F, Imai Y, Maeda A, Tsukada I, Komiya S and Gonnelli R S 2014 {\em Supercond. Sci. Technol.} {\bf 27} 124014

\bibitem{TamegaiSUST}
Sun Y, Taen T, Yamada T, Tsuchiya Y, Pyon S and Tamegai T 2015 {\em Supercond. Sci. Technol.} {\bf 28} 044002

\bibitem{TanatarSUST}
Cho K, Tanatar M A, Ni N and Prozorov R 2014 {\em Supercond. Sci. Technol.} {\bf 27} 104006

\bibitem{Capone2014}
de' Medici L, Giovannetti G and Capone M 2014 {\em Phys. Rev. Lett.} {\bf 112} 177001

\bibitem{Ding2008} 
Ding H {\em et al.} 2008 {\em Europhys. Lett.} {\bf 83} 47001

\bibitem{Okazaki}
Okazaki K {\em et al.} 2014  {\em Sci. Rep.} {\bf 4} 4109

\bibitem{Borisenko2012} 
Borisenko S V {\em et al.} 2012 {\em Symmetry} {\bf 4} 251 

\bibitem{BianconiNP2013} 
Bianconi A 2013 {\em Nature Phys.} {\bf 9} 536

\bibitem{PeraliSUST}
Guidini A and Perali A 2014
{\em Supercond. Sci. Technol.} {\bf 27} 124002

\bibitem{Kanigel} 
Lubashevsky Y, Lahoud E, Chashka E, Podolsky E and Kanigel A 2012 {\em Nat. Phys.} {\bf 8} 309

\bibitem{Perali2000} 
Perali A, Castellani C, Di Castro C, Grilli M, Piegari E and Varlamov A A 2000 {\em Phys. Rev. B} {\bf 62} 9295 

\bibitem{BenfattoSUST} 
Fanfarillo L and Benfatto L 2014 
{\em Supercond. Sci. Technol.} {\bf 27} 124009

\bibitem{VarlamovSUST}
Koshelev A E and Varlamov A A 2014 {\em Supercond. Sci. Technol.} {\bf 27} 124001

\bibitem{Komendova2012}
Komendov\'{a} L, Chen Y, Shanenko A A, Milo\v{s}evi\'{c} M V and Peeters F M 2012 \textit{Phys. Rev. Lett.} \textbf{108}  207002 

\bibitem{ArkadySUST}
Shanenko A A {\em et al.} 
2015 {\em Supercond. Sci. Technol.} {\bf 28} 054001

\bibitem{Blatt} Blatt J M and Thompson C J 1963 {\em Phys. Rev. Lett.} {\bf 10} 332

\bibitem{yam}
Yamada M {\em et al.} 2013 {\em Phys. Rev. Lett.} {\bf 110} 237001

\bibitem{boz}
Bollinger A T {\em et al.} 2011 {\em Nature} {\bf 472} 458

\bibitem{Perali96} 
Perali A, Bianconi A, Lanzara A and Saini N L 1996  {\em Solid State Comm.} {\bf 100} 181

\bibitem{per2012}
Perali A, Innocenti D, Valletta A, Bianconi A 2012 {\em Supercond. Sci. and Technol.} {\bf 25} 124002

\bibitem{Shanenko2006}
Shanenko A A, Croitoru M D, Zgirski M, Peeters F M and Arutyunov K 2006 \textit{Phys. Rev. B} \textbf{74} 052502, and references therein.

\bibitem{Shanenko2012} 
Shanenko A A, Croitoru M D, Vagov A V, Axt V M, Perali A and Peeters F M 2012 {\em Phys. Rev. A} {\bf 86} 033612 

\bibitem{Mizohata} 
Mizohata Y, Ichioka M and Machida K 2013 {\em Phys. Rev. B} {\bf 87} 014505 

\bibitem{MultiSuper} International Network on Multi-component Superconductvity and Superfluidity: http://www.multisuper.org 

\bibitem{AlbinoSUST}
Portela F S, Corredor L T, Barrozo P, Jung S-G, Zhang G, Vanacken J, Moshchalkov V V and Aguiar J A 2015 {\em Supercond. Sci. Technol.} {\bf 28} 034001

\bibitem{CapraraSUST}
Caprara S, Bucheli D, Scopigno N, Bergeal N, Biscaras J, Hurand S, Lesueur J and Grilli M 2015 
{\em Supercond. Sci. Technol.} {\bf 28} 014002

\bibitem{MauroSUST}
Doria M M, Vargas-Paredes A A and Cariglia M 2014 
{\em Supercond. Sci. Technol.} {\bf 27} 124008

\bibitem{BianconiSUST}
Bianconi A, Poccia N, Sboychakov A O, Rakhmanov A L and Kugel K I 2015 {\em Supercond. Sci. Technol.} {\bf 28} 024005
  
\end{thebibliography}
\end{document}